# DUAL CHAMBER LASER ION SOURCE AT LISOL

Yu. Kudryavtsev[1], T.E. Cocolios[1], J. Gentens[1], M. Huyse[1], O. Ivanov[1], D. Pauwels[1], T. Sonoda[1,2], P. Van den Bergh[1], P. Van Duppen[1]

[1]Instituut voor Kern- en Stralingsfysica, K.U.Leuven, Celestijnenlaan 200D, B-3001 Leuven, Belgium

[2]RIKEN, 2-1 Hirosawa, Wako, Saitama 351-0198, Japan

**Abstract**

A new type of the gas cell for the resonance ionization laser ion source at the Leuven Isotope Separator On Line (LISOL) has been developed and tested at off-line and on-line conditions. Two-step selective laser ionization is applied to produce purified beams of radioactive isotopes. The selectivity of the ion source has been increased by more than one order of magnitude by separation of the stopping and laser ionization regions. This allows to use electrical fields for further ion purification.

Corresponding author: Kudryavtsev Yu., Instituut voor Kern- en Stralingsfysika, K.U.Leuven, Celestijnenlaan 200 D, B-3001 Leuven, Belgium.

Tel: ++32-16-327271; Fax: ++32-16-327985;

E-mail: yuri.kudryavtsev@fys.kuleuven.be

## 1. Introduction

For the production of short-lived radioactive ion beams often high-pressure noble gases are used as a stopping media for the nuclear reaction products: the ion guide technique that was pioneered in the early 1980s at the University of Jyväskylä [1,2] and the ion catcher technique implemented recently at in-flight radioactive beam facilities [3-7]. In both methods one tries to avoid the recombination of the stopped recoil ions. In the ion guide method, this is achieved by minimizing the stopping volume in the vicinity of the cell exit hole and by fast extraction of ions by the gas flow. In the gas catcher method, which is mainly designed for in-flight separators, DC and RF electrical fields are used to prevent the ions from touching the walls of the gas cell and to move them towards the exit hole where the motion is taken over by the gas flow. Furthermore, the electrical fields separate the ions and electrons created during the stopping process and reduce the ion recombination. The high intensity of the incoming beam can cause space charge effects and reduce the efficiency [8,9]. Because of its small recombination coefficient, only helium is used.

Opposite to the above mentioned methods, the operational principle of the laser ion source is based on an element-selective resonance multi-step laser ionization of neutral atoms that after production in a nuclear reaction are thermalized and neutralized in a buffer gas where a weakly ionized plasma is created by the primary beam, the recoil ions and the radioactivity. This method was developed at K.U.Leuven in the early 1990s [10-12], and is used since then at the Leuven Isotope Separator On Line (LISOL) facility (Belgium) to produce short-lived radioactive isotopes. Recently, it has been implemented at the IGISOL facility (Finland) [13,14]. For completeness, we note that the laser ionization spectroscopy in a gas cell has been developed as well [15]. Depending on the nuclear reaction different types of gas cells have been used for proton-induced fission and light- and heavy-ion induced fusion evaporation. In these experiments helium or argon at 500 mbar pressure as a buffer gas are typically used, however argon is preferentially used because of the larger recombination coefficient and the larger stopping power for energetic recoils. In both fission and fusion gas cells the stopping/thermalizing and laser ionization zones are not separated physically [16,17]: stopping and laser ionization happens in the same gas cell. As a result the primary accelerator beam, recoils and radioactivity influence the plasma conditions in the laser ionization zone and the ions created by resonant laser ionization (laser/photo ions) at a distance of a few cm from the primary beam path recombine fast (on a ms time scale), see e.g. Fig. 4 in [17]. Furthermore, unwanted ions are present and create an isobaric background for the experiments [18]. To reduce the recombination of laser ions, the survival of non-resonantly produced ions and the creation of unwanted ions by different processes (see further), a pulsed beam mode in anti-phase with the mass separator time gate was used. In this way the mass separated ion beam was only transported to the detection set-up when the primary beam was not present [12].

There are several sources of non-selective ionization in the laser-ionization zone; these are ion scattering off the primary beam creating a flux of energetic ions in the laser ionization zone and hard UV and X-ray radiation from the target/window material and from the buffer gas. The nuclear reaction products, especially in the case of fission, can also contribute to the weakly ionized plasma and unwanted ion creation in this zone. When the primary beam passes through the gas it transfers most of its energy to atoms via inelastic collisions. This energy is dissipated in the gas via the emission of δ-electrons and photons. The δ-electrons with energy up to a few keV have a short range in the gas (less than 1mm) but they cause excitation of inner electrons of the buffer gas atoms. This excitation results in vacancy cascades with emission of Auger electrons and fluorescent photons. The keV energy photons can initiate further excitation and ionization [19,20]. The probability of emitting fluorescent photons is higher for heavier atoms. In case of argon gas about 12 % of the deposited energy in the excitation of K-shell electrons goes to fluorescence and the rest to Auger electron emission [21]. The gas cell windows, made of molybdenum, give 76% of excitation energy in the X-ray region. The vacancies in the K and L shells cause also an electron-shake off process that leads to the creation of multi-charged ions (with maximum 4+ state for argon) [22] that emit hard UV radiation. Most of the energy is deposited in the beam path but the scattered ions, the photons and energetic reaction products ionize the gas at larger distance from the point of initial ionization and a low-density plasma is created far from the beam path. This causes recombination of laser-produced ions, thus reducing the ion source efficiency, and the creation of unwanted ions. Collection of the unwanted ions prior to laser ionization using electrical fields is prevented due to the space-charge effect present in the gas cell [8,17].

In this article we present a new gas cell with separated stopping- and the laser-ionization chambers. In this design the laser ionization zone is not in direct view from the accelerator-beam path and the trajectories of recoils. This should allow us to avoid recombination of laser-produced ions, to use the accelerator beam in DC mode and to collect not-neutralized ions before laser ionization using electrical fields.

## 2. Experimental setup

### 2.1. Dual chamber gas cell

The dual chamber gas cell for proton-induced fission is shown on Fig. 1. It consists of stopping and ionization chambers that are connected via an elbow channel. The stopped recoils are brought from the stopping volume to the laser ionization volume by the gas flow.

The noble gas, purified down to the ppb level in a getter-based purifier, enters the gas cell via the ring slit that homogeneously distribute the gas across the cell. The inner diameter of the stopping chamber is 4 cm and its length is 6 cm. The accelerator beam enters the cell through a molybdenum foil of 4 µm in thickness. The target is installed on the tilted surface of the insert that is fixed in the stopping chamber. The angle between the target surface and the incoming accelerator beam can be changed. This angle equals to 16 degrees for proton-induced fission of uranium-238 and 35 degrees for heavy-ion induced fusion evaporation reactions. The shape of the insert guarantees a turbulent free homogeneous gas flow towards the elbow. This is confirmed by gas flow simulations as presented in paragraph 3.2.

The laser beams enter the ionization chamber (30 mm long and 10 mm in diameter) longitudinally through a quartz window and ionize atoms along the chamber axis, Fig. 1. This laser beam path can be used to monitor the ion behavior in the ionization chamber. Atoms of stable nickel or cobalt isotopes can be produced inside the gas cell by resistive heating of corresponding filaments. An additional extension of 12 mm in length allows transverse laser beam entrance near the exit hole region. In this case an ion collector, located upstream, can be used to collect non-neutralized ions that come from the stopping chamber without collecting the laser-produced ions. The ion collector plates are shaped according to ring electrodes with an inner diameter of 11 mm. The evacuation time of the laser ionized volume (both longitudinal and transverse) at the exit hole diameter of 0.5 mm is bigger than the time between two subsequent laser pulses of 5 ms guaranteeing that all atoms have been irradiated by laser light. Ions leaving the gas cell are captured by a Sexupole Ion Guide (SPIG) and transported towards the mass separator.

## 2.2. Sextupole ion guide

The ions coming out of the cell have essentially the jet velocity of the carrier gas. The RF voltage applied to the SPIG rods provides radial confinement of the ions. A DC voltage up to 300 V of either polarity (+ or -) can be applied between the gas cell and SPIG rods. In normal running conditions the SPIG rods are negatively based relative to the gas cell. In this case molecular ions that can be formed inside the gas cell after laser ionization are dissociated if the voltage is large enough [23,24]. In the case of a positive polarity, the ions from the gas cell are repelled. However in the longitudinal ionization mode part of the laser beam intensity goes through the exit hole and can ionize atoms outside the gas cell; only those ions created inside the SPIG are then transported towards the mass separator. This is the so-called LIST mode (Laser Ion Source Trap) proposed for a hot cavity in [25]. The results of our studies on the LIST mode, coupling a gas cell with an RF-ion guide, combined with laser ionization in the RF structure and showing the possibility to do laser spectroscopy free of pressure broadening are presented in a separate paper [26].

**2.3. Detection**

After mass separation, the radioactive ions are implanted into the movable tape of a tape station. Two high-purity germanium gamma detectors and three scintillation beta detectors surround the implantation point. Stable ions are detected by a secondary electron multiplier. An ion counting system and a digital oscilloscope were used to acquire ion time profiles after laser ionization. More details can be found in [27, 24].

**2.4. Laser system**

Two-step two-color schemes are used for the resonance laser ionization of stable and radioactive atoms. The first step laser excites atoms in an intermediate state followed by a transition into an autoionizing state by the second step laser. The laser system consists of two dye lasers pumped by two excimer XeCl lasers with a maximum pulse repetition rate of 200 Hz [12]. The first step laser radiation is frequency doubled in a second harmonic generator. The laser beams of the first and second steps are overlapped at very small angle in the ionization chamber of the gas cell located 15 m away from the laser system. The diameters of laser beams are about 4-6 mm. The laser pulse widths and bandwidths equal to 15 ns and 0.15 cm$^{-1}$, respectively.

**3. Evacuation properties of the gas cell.**

Important parameters influencing the efficiency of the gas cell are its evacuation time and diffusion losses towards the walls of the atoms thermalized in the buffer gas. Those were studied by measuring the ion time profiles and by calculation of the flow pattern and trajectories of atoms and ions in the cell.

**3.1 Ion time profiles**

Fig. 2 shows the laser ion time profiles of stable cobalt ions produced in the ionization chamber without extension, longitudinally, using helium and argon as buffer gas. The laser ionization of the continuous flow of cobalt atoms takes place at t = 0. The evacuation time from the elbow region in the case of argon equals about 35 ms, which is 3.5 times longer than in the case of helium, 10 ms. This ratio reflects the difference in the conductance of the exit hole of 0.5 mm in diameter for argon and helium, 35 cm$^3$/s and 112 cm$^3$/s, respectively. The ion time distribution is a measure of the spatial distribution of the cobalt atoms in the laser beam path at the moment of laser ionization. The bump in the argon time profile at 35 ms (10 ms in helium) reflects a higher cobalt atom density in the elbow region, see paragraph 3.2.

The information about the evacuation time from the stopping chamber of the gas cell can be obtained if atoms are injected into the cell during a short time. We applied the same technique as explained in ref. [8,17] using an accelerated 185 MeV $^{58}$Ni beam that is stopped in the gas cell, evacuated to the laser ionization zone, resonantly ionized, mass-separated and detected. Fig. 3a shows the ion time profile of nickel atoms from the cell at different argon pressure after injection of a 50 ms long pulse of 185 MeV $^{58}$Ni beam at t = 0 with a beam intensity of 0.25 pnA measured in a DC mode. In this measurement the insert was not inside the stopping chamber and the lasers were running at 100 Hz. At 480 mbar, the nickel beam is stopped approximately in the center of the cell, 30 mm from the entrance window and the maximum of the mass-separated ion signal is observed at 320 ms. Note that the ion signal is resonant and only $^{58}$Ni ions are present on mass 58. If the pressure is reduced, the beam is stopped further away from the entrance window and the evacuation time and diffusion losses increase. At 400 mbar 92% (relative to 480 mbar) of the ions are extracted and mass-separated. At 300 mbar the maximum of the ion signal is at 600 ms and only 57% of the ions are extracted and mass-separated compared to 480 mbar. If the pressure is reduced further down to 200 mbar the ion signal drops to 11% compared to 480 mbar (not shown in Fig. 3a). Fig. 3b shows the calculated evacuation time profiles at 480, 400 and 300 mbar of argon that will be discussed in the next paragraph. Interesting to note that the evacuation time profile of the standard LISOL gas cell (Fig. 3c) is longer than in the dual chamber gas cell as the extra delay from the ionization chamber is compensated by a better match between the effective stopping volume and the gas flow.

**3.2. Gas flow simulation**

The gas flow simulations were performed by using the COSMOS-Floworks 2006 program. It takes into account the exact dimensions of the gas cell including the ring slit for the gas entrance, the target holder and the ion collector. Fig. 4 shows gas flow trajectories in the cell including the fission target and with the extension for the transverse laser ionization. The flow is laminar without turbulences. The elbow causes the flow lines in the ionization chamber to be closer to the left-hand side of the cell. As a consequence, the overlap of the laser beams (diameter 4-6 mm) with the flow of atoms along the ionization chamber is not complete. The good overlap with the flow of atoms in the elbow region explains the bump at longer times in the shape of the cobalt ion time profiles (at 10 ms and 35 ms for helium and argon, respectively) as shown in Fig. 2.

The evacuation time and the diffusion losses of nuclear-reaction products were calculated using the real target geometry. The initial fission products distribution in the cell was calculated for fission recoils from the 10 μm uranium-238 target in 500 mbar argon. The trajectories of 1192 fragments were calculated using a macroscopic simulation including diffusion losses. Fig 5a shows the simulated evacuation time profile of all recoils that survive the diffusion and arrive to the exit hole. A total of 394 atoms were found at the exit hole resulting in a transport efficiency of 33 %. The first atoms arrive after 60 ms and within 600 ms most of the atoms are evacuated from the cell. For comparison, in Fig. 5b the time distribution of the fission products escaping from the standard fission gas cell [16] is shown. Note the presence of atoms at very short times, which is due to the fact that in the standard cell part of the fission recoils are stopped very close to the exit hole. The diffusion losses in this cell are less in comparison to the one in the shadow cell. The simulation shows that 40.8% of the recoils stopped in the gas cell after fission are transported to the exit hole.

The delay in evacuation of the recoils from the gas cell can cause an additional reduction of the total ion source efficiency due to radioactive decay inside the cell during the transport to the exit hole. This effect was calculated as a function of the half-life time of the studied isotope for the dual chamber cell and for the standard fission cell for an exit hole diameter of 0.5 mm and of 1 mm. The results are shown in Fig. 6. It is obvious that the survival efficiency of the both cells is larger in case of a 1 mm exit hole compared to 0.5 mm because of a faster gas flow through the exit hole. The dual chamber cell with a 0.5 mm exit hole has a lower efficiency for isotopes with half-lives less than 1s. This is again related to the delay time in the elbow region and ionization chamber. However this delay is reduced by a factor of four by increasing the exit hole diameter up to 1 mm; the efficiency of the shadow cell is then larger compared to that of the standard fission cell for isotopes with half-lies larger than about 40 ms.

The gas flow simulation allows to explain reasonably well the experimental results. In paragraph 3.1, Fig. 3a and 3b show experimental and calculated time profiles of the evacuated nickel ions after the pulsed injection of the 185 MeV $^{58}$Ni beam into the gas cell at different argon pressures. The initial position of the stopped ions was calculated using the SRIM code. Then the trajectory of each ion was calculated using the flow data. The evacuation time is the time elapsed between the creation of the ion and the moment of successful arrival of this ion at the exit hole. The injection time of 50 ms was taken into account. The calculated efficiency, defined as the number of nickel atoms/ions transported to the exit hole versus the number of incoming nickel ions, equals 68%, 61%, 49% and 14% at 480 mbar, 400 mbar, 300 mbar and 200 mbar, respectively. The increasing loss with decreasing pressure is due to diffusion to the walls of the gas cell but the strong reduction between 300 and 200 mbar is mainly due to incomplete stopping in the gas. The relative efficiency at 400 mbar, 300 mbar and 200 mbar, relative to the one at 480 mbar equals 90%, 72%, and 21% respectively. These values can be compared with the experimental values of 92%, 57% and 11% presented in paragraph 3.1 (see Fig. 3).

## 4. Laser ionization

The concept of the dual chamber gas cell was investigated by using longitudinal and transverse laser ionization of stable atoms evaporated from a filament in off-line and on-line conditions as well as radioactive isotopes produced in fusion evaporation and fission reactions.

### 4.1. Longitudinal laser ionization

*4.1.1. Off-line test*

An important element of the gas cell is the ion collector (IC) (see Fig.1). Its performances were tested off-line by longitudinal laser ionization of stable nickel atoms evaporated from the filament. Time profiles of the mass-separated nickel ions at different voltages applied to the ion collector are shown in Fig. 7. The voltage pulses (5 ms long) of different polarities but equal amplitude are applied to the opposite electrodes 10 ms after the laser pulse. This measurement is performed in the cell without extension. If the amplitude of the pulse is more than 24 V, essentially all ions in the time interval between 15 - 33 ms are collected (the dynamic range of this measurement is about 200). This time interval corresponds to ions located in the IC region when the voltage pulse was applied. When the IC pulse is made 10 ms longer the ions produced in the elbow region are also collected. In case of a DC voltage on the ion collector, only ions produced very close to the exit hole survive the collection because of the weak electrical field in this region.

*4.1.2. On-line test*

The performance of the ion collector was also tested in the presence of a 1 eμA 265 MeV $^{40}Ar^{+11}$ beam in the cell with extension at 500 mbar of argon as the buffer gas. In this case, in the stopping volume, about 3e17 ion-electron pairs·$s^{-1}$·$cm^{-3}$ are created in the cyclotron beam path, see equation 1 and table 1 in ref. [17], resulting in a plasma density of about 5e11 ions·cm$^{-3}$ in this region. This is extremely high for applying any electrical field for the ion collection [8]. However in the laser ionization chamber, the plasma conditions are completely different and the ion collector can be used for purification. Fig. 8 shows time profiles of stable nickel atoms after longitudinal laser ionization in 500 mbar of argon with a laser repetition rate of 5 Hz (a laser pulse every 200 ms). The length of the ion signal without cyclotron beam (the end of the signal is defined when the flat part in the range 30 - 50 ms drops by a factor of 2) equals to 76 ms, which corresponds to the evacuation time of ions from the ionization chamber with extension. If the cyclotron beam is switched ON the length of the signal is shorter (~63 ms). The reason for the shorter pulse is the neutralization of the laser-produced ions in the elbow region, which are in a direct view from the cyclotron beam path. The neutralization is due to the processes discussed in the introduction. Note that the amplitude of the ion signal from the rest of the ionization chamber stays almost the same indicating that the shielding effect indeed works. At longer times (>100 ms), the signal does not decrease further but saturates and even crosses the curve of laser ions without cyclotron beam, indicating their beam-related non-resonant character. If the ion collector is switched ON (applied DC voltage of 40V) the nickel ion signal dramatically decreases after 15 ms and drops by three orders of magnitude at 60 ms. It is important to note that the ion time profile at $0 < t < 15$ ms is only weakly influenced by the presence of the cyclotron beam and the IC.

**4.2. Transverse laser ionization**

When using transverse laser ionization the cyclotron beam and ion collector can run in DC mode. In this case the laser pulse repetition rate should be high enough: the evacuation time of the laser-irradiated volume should be more than the time between two subsequent laser pulses. The time profiles of transversely laser produced nickel ions at different laser repetition rates are shown in Fig. 9. The lasers are triggered at t = 1 ms. At low pulse repetition rate of 20 Hz one observes a time profile with a full width at half maximum of 5.5 ms decreasing to the noise level before the next laser pulse is fired. If increasing the repetition rate to 100 Hz, a pulse structure is still present indicating that not all atoms have been ionized. At 200 Hz saturation is almost reached. This is supported by Fig. 10 where the ion count rate as a function of the laser pulse repetition rate is shown for transverse laser ionization. The time profiles were taken with the ion collector ON and OFF and no influence of the IC voltage was observed. This is in agreement with the time profile of the ion signal with longitudinal laser ionization and ion collector ON (Fig. 8), where the nickel ions are not collected during the first 15 ms.

Fig. 10 shows also the ion count rate in function of the laser pulse repetition rate for the longitudinal laser ionization with ion collector ON. Since the signal with longitudinal ionization (20 ms) is longer than with transverse one (5.5 ms), the saturation on Fig. 10 is observed at lower (50 Hz) repetition rate. However the saturation level is less because the ions produced in the upstream region of the ionization chamber are collected. The presence of the cyclotron beam does not change the time profiles and the saturation curves. The influence of the ion collector on the laser selectivity of stable $^{58}$Ni in the presence of the 265 MeV $^{40}$Ar$^{+11}$ beam was tested with lasers running at 200 Hz. The laser selectivity, defined as the ratio of the $^{58}$Ni count rate with lasers ON to the count rate with lasers OFF, equals to 155 without the IC voltage and increases up to 7500 at the IC voltage of 40V. This shows that the ion collector can be used to improve the selectivity of the laser ion source for radioactive isotopes and that the dual camber gas cell can be used under DC primary beam conditions.

*4.2.1. Heavy ion-induced fusion-evaporation reaction*

The dual chamber gas cell was tested in on-line conditions using radioactive $^{94}$Rh isotopes produced by impinging a $^{40}$Ar beam on a $^{58}$Ni target. Fig. 11a shows a β-gated γ spectrum with transverse lasers tuned in resonance to rhodium when the ion collector is ON. Only rhodium γ lines are present; the total $^{94}$Rh yield equals 12600 (850) at/μC. A similar spectrum and yield of 13600 (650) at/μC are observed if the IC is OFF. Yields of 14670 (100) at/μC and 14690 (100) at/μC have been deduced from counting the number of β particles without IC and with IC, respectively. The slightly larger yields obtained in the case of β counting is due to contributions from the decay long-lived daughter isotopes. As no significant difference in yield with and without IC is observed it can be concluded that the ion collector does not collect laser-ionized radioactive rhodium isotopes produced in heavy-ion fusion reaction. Fig. 11b shows the spectrum on mass 94 when the lasers are OFF and IC is OFF. Very weak γ lines belonging to $^{94}$Rh are present in the spectrum since some ions survive neutralization and reach the exit hole. A laser selectivity for $^{94}$Rh of 500 could be deduced. The selectivity for this radioactive isotope is three times larger than for the stable nickel isotope on mass 58, see paragraph 4.2. This can be explained by the presence of background molecular ions at mass 58. If the IC voltage is applied no γ lines are observed, Fig. 11c. The total selectivity has been determined from counting the number of β particles and was more than 2200. This increase in selectivity of the dual chamber gas cell ion source opens new possibilities to perform spectroscopy studies of neutron-deficient isotopes in the N=Z region.

*4.2.2. Proton-induced fission reaction*

A striking difference in selectivity of the standard laser ion source for different β⁻-decaying states within the same isotope, produced in a proton-induced fission of uranium was observed. The laser selectivity changed from 200 for $^{112m}$Rh to 3-4 for $^{112g}$Rh [28]. This was explained as due to different ways the β⁻-decaying states were populated; directly in the fission reaction or through β⁻ decay. In the dual chamber gas cell, the ion collector can be used as an additional tool to understand the different selectivity for high- and low-spin isomers. Radioactive neutron-rich $^{112}$Rh isotopes were laser ionized in the same way as neutron-deficient $^{94}$Rh isotopes described in the previous paragraph. Fig. 12a shows a β-gated γ spectrum with transverse lasers tuned in resonance to rhodium atoms when the ion collector is OFF. As in the case of $^{94}$Rh, the ion collector mode (ON or OFF) has almost no influence on the laser-produced ions. Transitions only present in the decay of the high spin isomer are denoted by m while the ones present in both decays are denoted by m+g. In the present set-up, laser radiation ionizes the ground- or metastable nuclear isomer state with equal efficiency as the isomer shift is much smaller compared to the total laser line width. The half-lives of both isomers ($T_{1/2}(^{112m}Rh) = 6.8$ s $T_{1/2}(^{112g}Rh) = 2.1$ s), are much longer than the evacuation time of the gas cell, so the decay losses inside the cell can be neglected. The insert in Fig. 12a shows a simplified decay scheme of the mass 112 chain. Next to direct feeding in the fission reaction, also feeding through the β⁻-decaying parent nucleus ($^{112}$Ru, $T_{1/2} = 1.75$ s) can occur and only the low-spin ground state of $^{112g}$Rh receives feeding from the even-even mass $^{112}$Ru (I=0$^+$) nucleus. Fig. 12b shows the spectrum accumulated at mass 112 when the lasers are OFF and the IC is OFF. The γ lines belonging to the high-spin isomer ($^{112m}$Rh) are reduced by a factor of 25 while γ lines belonging to the decay of $^{112g}$Rh are only reduced by a factor of 2.3 (note that the 349 keV line intensity in Fig. 12a stems for 33% and 87% from the ground state and high-spin isomer, respectively). If the IC voltage is applied all γ lines of the high-spin isomer disappear, Fig. 12c, however the intensity of the lines fed by the $^{112g}$Rh is only slightly reduced. The selectivity for the $^{112m}$Rh is estimated to be above 1000 and the selectivity for $^{112g}$Rh is increased from 2.3 to 3.1.

The different behaviour of isotopes whether or not receiving feeding from parent nuclei can be explained by the fact that these parent nuclei stick to the inner surface of the gas cell and/or to the SPIG rods and subsequently decay. The majority of all fission products are neutralized in the stopping chamber and pass through the ion collector in the laser ionization chamber. Some of them are deposited in the exit hole region or on the SPIG rods instead of being pumped away. Their β decay can detach the daughter nucleus from these surfaces and leaves the nucleus in an ionized state, mostly in a 1+ state. These ions can then be further transported by the gas flow in the cell or outside the cell by a combination of the buffer gas and the electrical fields applied. From the experiments described below we can conclude that the main contribution to the non-resonant production of such daughter nuclei comes from the rods of the RF ion guide. Fig. 13a shows the calculated distribution of atoms from the gas jet that hit the SPIG rods for a distance between the cell and the rods of 1.5 mm. Most of the atoms are deposited in the beginning of the RF structure. If the SPIG is displaced further, the relative amount in the beginning is increased. Fig. 13b shows the calculated argon pressure along the central line of the SPIG for a gas cell pressure of 500 mbar and an exit hole diameter of 0.5 mm. The energy of the recoiling daughter ions after a $\beta^-$ decay depends on the Q-value. In the $^{112}$Ru - $^{112g}$Rh decay the $Q_\beta$-value is 3.4 MeV and the maximum recoil energy of $^{112g}$Rh ions is 72 eV, which corresponds to a range of 2 mm in argon at a pressure of 1 mbar. Also, the depth of the potential well created by the RF field of the SPIG is about 100 eV. Thus the recoiling $^{112}$Rh ions can be easily confined by the RF structure and then can get a longitudinal velocity due to collisions with argon atoms from the jet. The charge state distribution after $\beta^-$ decay for different isotopes displays a typical yield around 80% for the single-charge state and around 10% for the double-charge state [29] and makes it a very efficient ionization process. In order to measure the importance of the latter discussed process versus the deposition of mother nuclei inside the gas cell, a measurement of the mass separated yield of different isotopes were performed for a series of fission products as a function of the applied voltage between the exit hole and the SPIG. Fig. 14 shows the yields (with lasers OFF) of different isotopes as a function of the potential on the SPIG rods relative to the gas cell. By applying a positive potential (>30 V in case of argon), the ions created in the gas cell and between the cell and the rods are not transported through the SPIG [26]. For stable nickel ions created in the ionization chamber, the reduction factor at a SPIG potential of 42 V is more than 1000. The reduction factor for laser-produced $^{112m}$Rh was determined to be more than 100. However, for $^{112g}$Rh, the reduction is only 2 times, (Fig. 14). A similar small reduction of factor 2 is observed for $^{142}$Ba, which has $^{142}$Cs as a parent nucleus with a $Q_\beta$-value of 7.3 MeV and a lifetime of 1.7 s. These ions can only come from the $\beta^-$ decay of atoms sticking to the SPIG rods. A completely different situation is observed for the yield of $^{142}$Cs isotopes, which drops 55 times when applying 40 V. This can be explained by the fact that the mother nucleus for $^{142}$Cs is $^{142}$Xe, which is a gaseous element that does not stick to the SPIG rods. We observed the sticking effect in our previous work [24], where we measured the ion-source efficiency using long-lived radioactive $^{57}$Co evaporated from a resistively heated filament. About 30% of the evaporated atoms were found on the SPIG rods. The sticking of radioactive isotopes in atomic or molecular form to the RF structure and consequent decay leading to the production of unwanted isotopes can limit the selectivity of the LIST method for neutral-rich nuclei, coupled either to a gas cell or a hot cavity. If the energy of the recoiling ions is small, they can be captured in the radial direction by the RF field without stopping in a low-pressure gas. A way to reduce this effect is the reduction of the RF structure surface.

**5. Conclusions**

Results obtained with a new type of gas cell whereby the stopping volume of the nuclear reaction products including the primary beam path are separated from the laser ionization volume have been presented. In this dual chamber gas cell concept the direct ionization near the exit hole through hard X-rays is blocked and enables the use of electrical fields inside the gas cell. This leads to a strong increase of the selectivity. A laser selectivity of least 2200 has been achieved for exotic nuclei produced in fusion-evaporation reactions. However, for isotopes produced in fission reactions, which have strong feeding from the $\beta^-$-decaying mother nuclei, the selectivity is limited because of the deposition of radioactive mother atoms on the rods of the RF ion guide.

**Acknowledgments**

The authors wish to thank to the cyclotron group at CRC Louvain-La-Neuve for running the accelerator. This work was supported by FWO-Vlaanderen (Belgium), GOA/2004/03 (BOF-K.U.Leuven), the 'Interuniversity Attraction Poles Programme - Belgian State – Belgian Science Policy' (BriX network P6/23) and by the European Commission within the Sixth Framework Programme through I3-EURONS (Contract RII3-CT-2004-506065).

**Figures captures**

Fig. 1 A schematic drawing of the dual chamber laser ion source gas cell.

Fig. 2 Time profiles of cobalt ions after a single laser pulse for longitudinal ionization in the gas cell without extension with helium and argon (500 mbar) as buffer gas.

Fig. 3 The evacuation time profile of nickel atoms after injection of a 50 ms pulse of 185 MeV $^{58}$Ni beam at t = 0 ms, $I_{DC}$ = 0.25 pnA) in argon from: a) the dual chamber gas cell at different argon pressure and laser repetition rate 100 Hz, b) the calculated evacuation time profiles from the dual chamber gas cell at 480, 400 and 300 mbar, c) the standard LISOL gas cell at 500 mbar of argon and laser repetition rate 20 Hz [16].

Fig. 4 Gas flow simulation in the dual chamber cell with an insert for the fission target and with an extension for transverse laser ionization.

Fig. 5 The simulated evacuation time profile of fission products for the 0.5 mm exit hole from a) the dual chamber gas cell, b) the standard LISOL fission gas cell [16].

Fig. 6 Survival efficiency against radioactive decay losses as a function of the half life of the isotope for the dual chamber- and the standard LISOL fission gas cells for 0.5 mm and 1 mm exit hole.

Fig. 7 Time profiles of the mass-separated nickel ions after a single laser pulse for longitudinal ionization in the gas cell without extension for different amplitudes of electrical pulses (5 ms long) applied to the ion collector (IC) with delay of 10 ms. The dashed line shows the effect of an increase in pulse length from 5 to 10 ms at 50 V.

Fig. 8 Time profiles of nickel ions after a single laser pulse for longitudinal ionization in the gas cell with extension. Note that the curves are not normalized to each other. The exit hole diameter is 0.5 mm and the argon pressure is 500 mbar.

Fig. 9 Time profiles of nickel ions after transverse laser ionization at different laser repetition rates. The exit hole diameter is 0.5 mm and the argon pressure is 500 mbar.

Fig. 10 Ion count rate of stable nickel ions as a function of the laser repetition rate for transverse and longitudinal ionization in the chamber with extension.

Fig. 11 β-gated γ spectrum obtained at mass 94: a) with lasers tuned in resonance to rhodium isotopes and IC - ON, b) Lasers - OFF and IC - OFF, c) Lasers - OFF and IC- ON. The measuring time is 300 s. The $^{94}$Rh atoms were produced in the $^{40}$Ar + $^{58}$Ni heavy ion fusion-evaporation reaction.

Fig. 12 β-gated γ spectrum on mass 112: a) with lasers tuned in resonance to rhodium isotopes and IC - OFF, b) Lasers - OFF and IC- OFF, c) Lasers - OFF and IC- ON, insert - decay chain at mass A = 112. The $^{112}$Rh atoms were produced in the proton-induced fission of $^{238}$U.

Fig. 13 a) Calculated distribution of atoms hitting the SPIG rods from an argon gas jet; b) calculated argon gas pressure on the axis of the SPIG for an exit hole diameter of 0.5 mm and a gas cell pressure of 500 mbar.

Fig. 14 Yield of $^{112}$Rh, $^{142}$Ba and $^{142}$Cs isotopes produced in the proton-induced fission of $^{238}$U as a function of the SPIG rods potential relative to the gas cell.

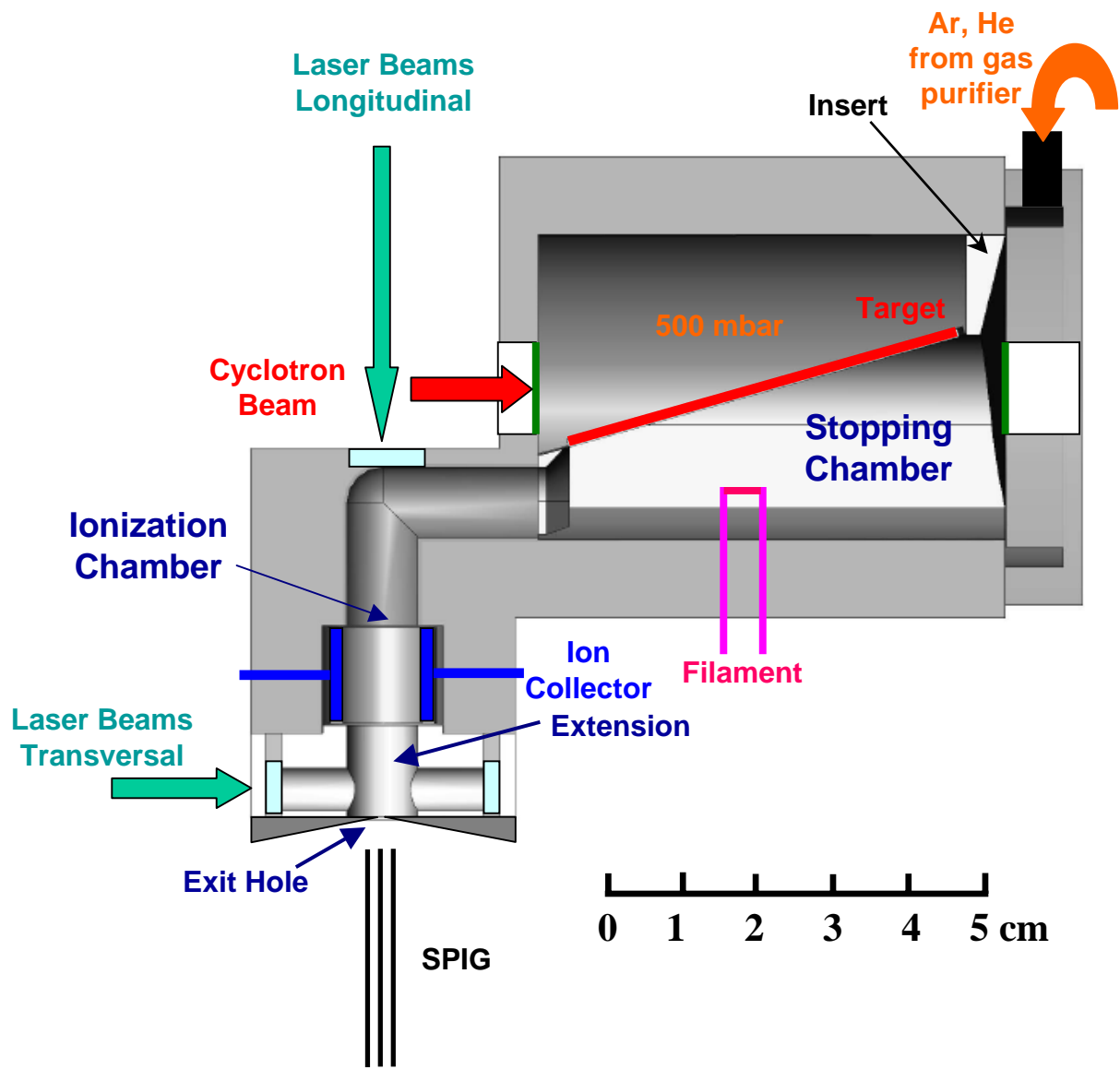

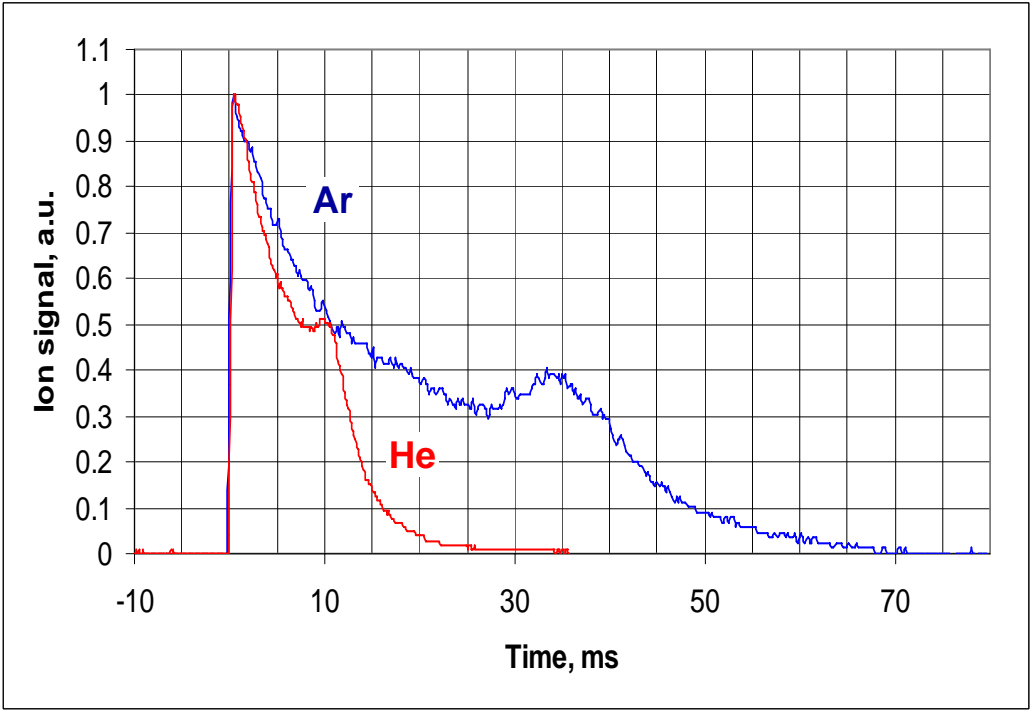

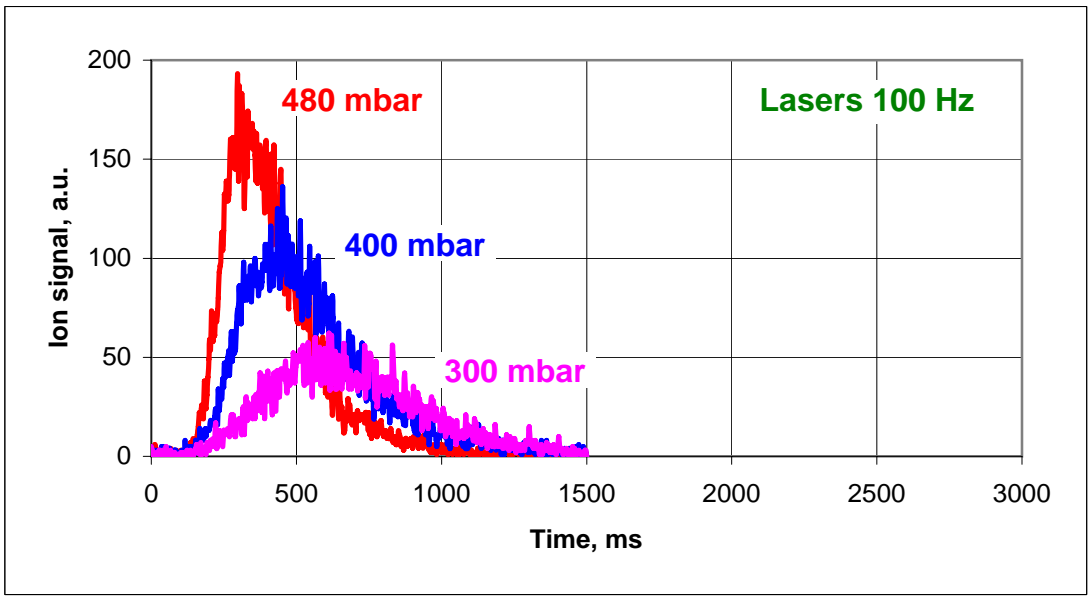
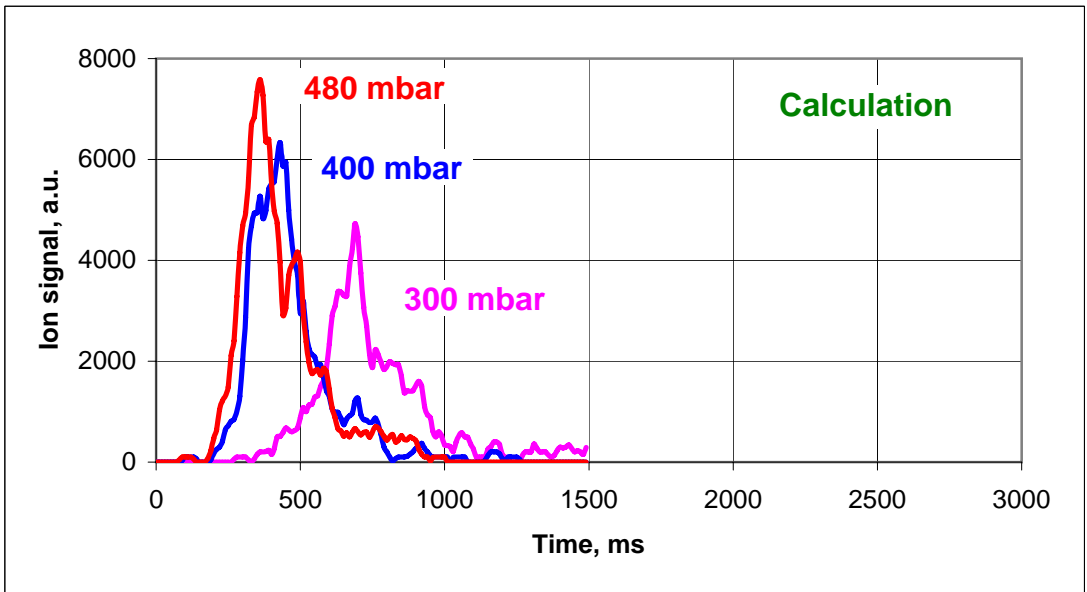
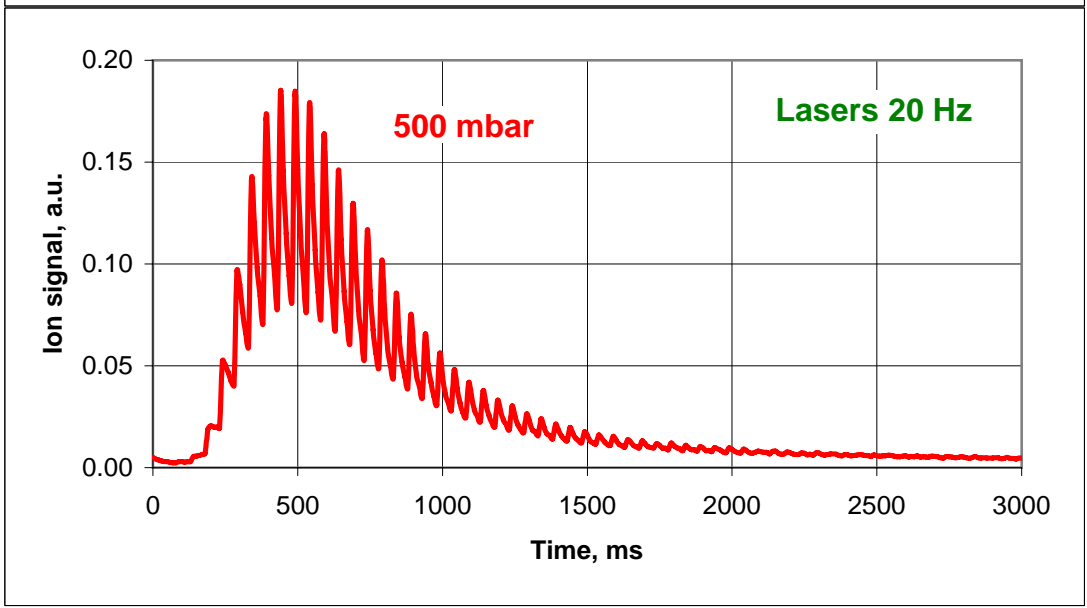

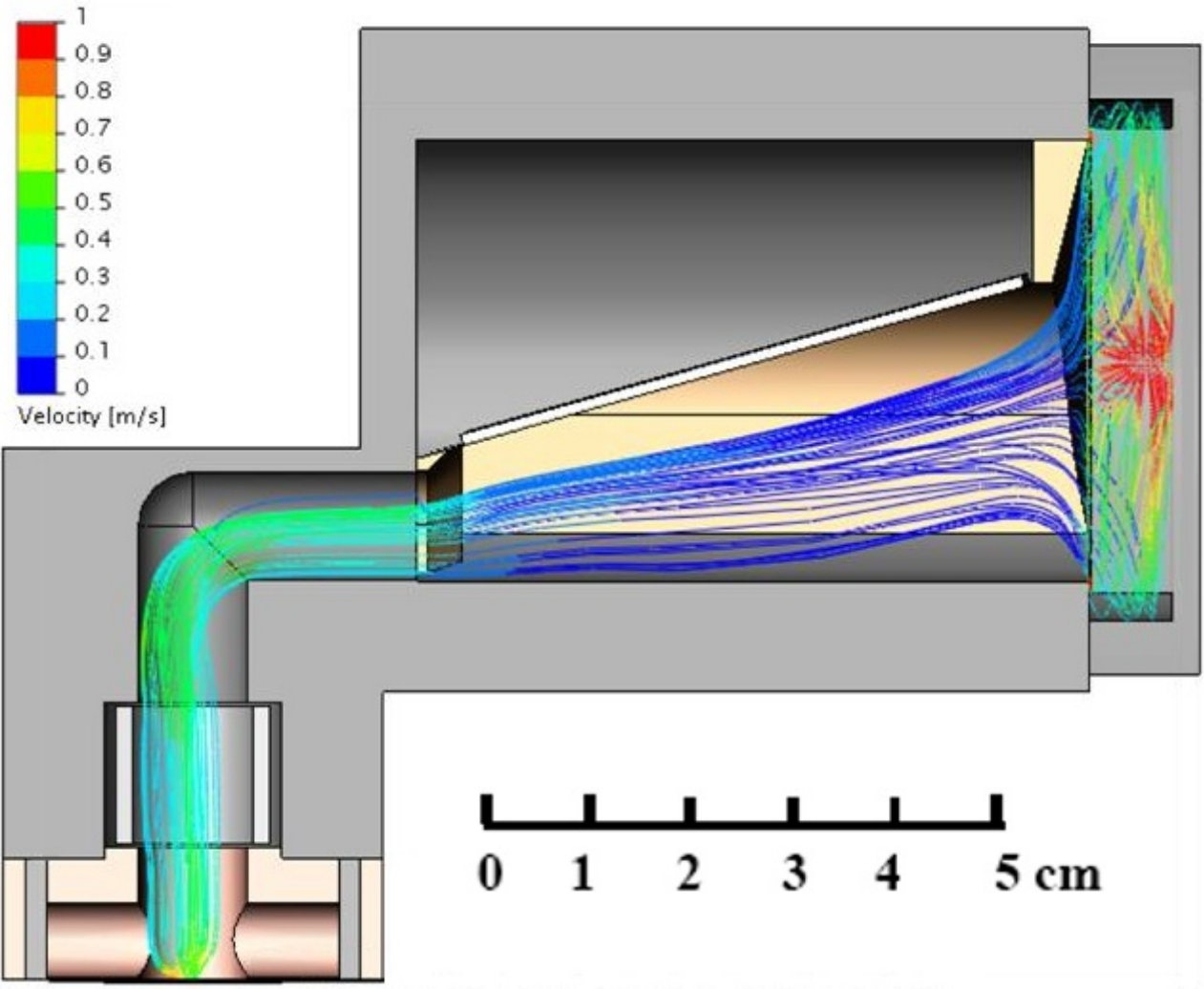

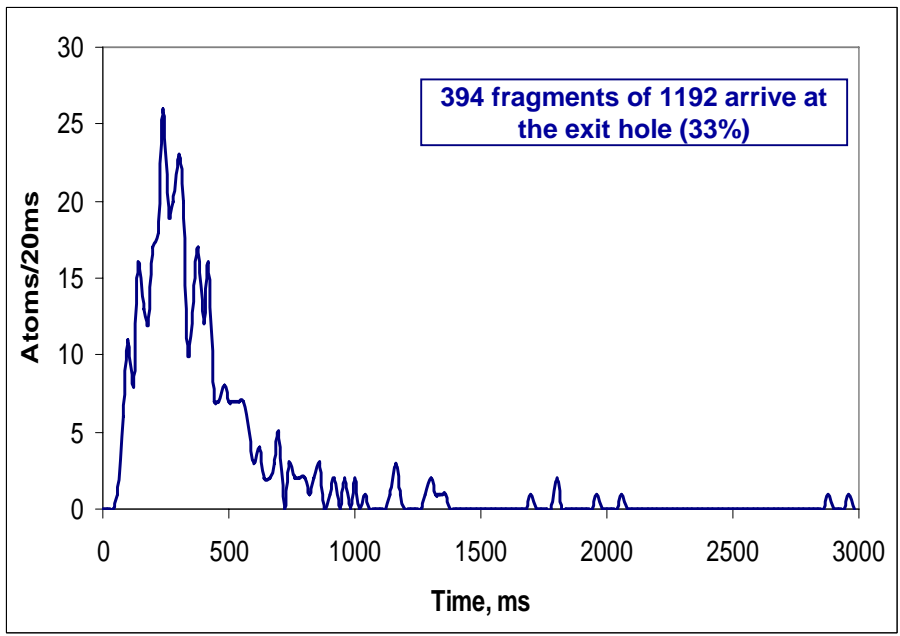

a)

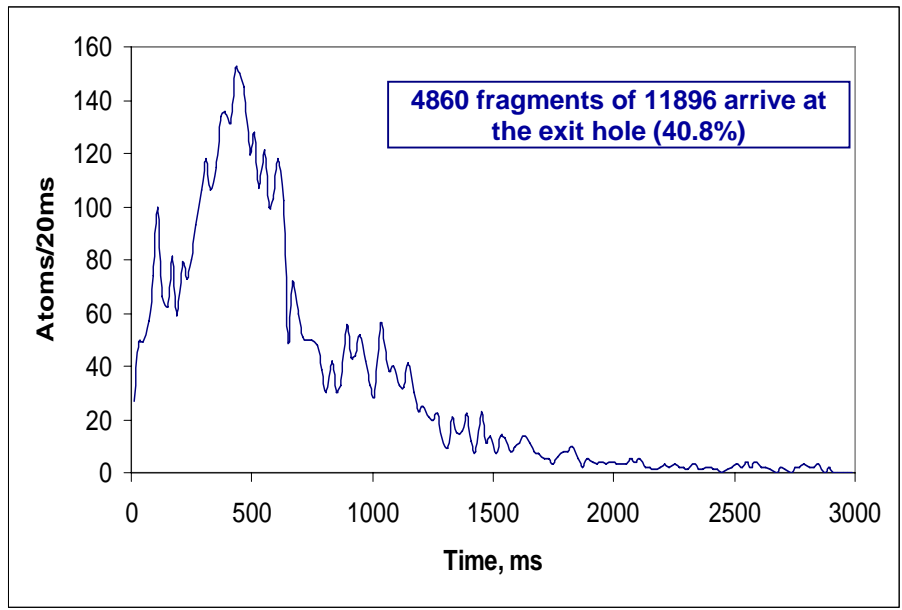

b)

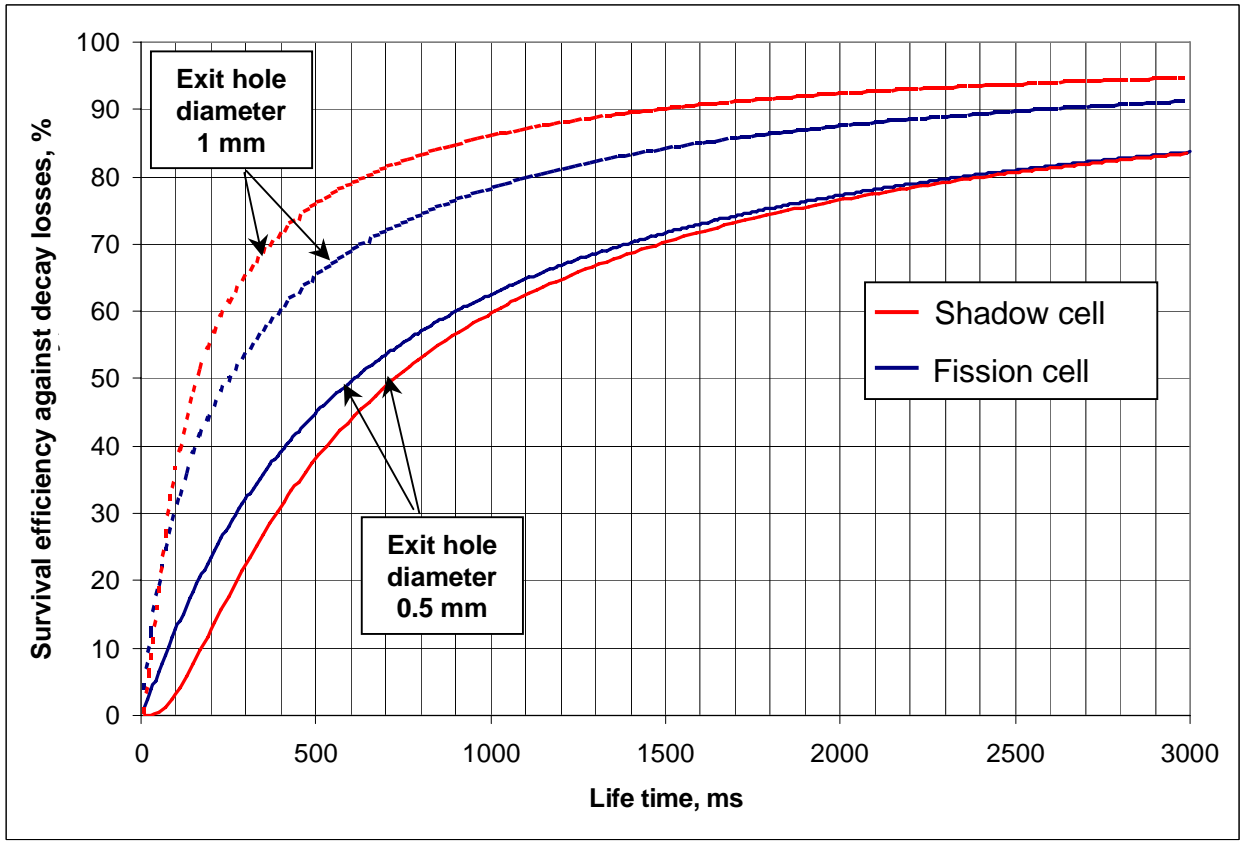

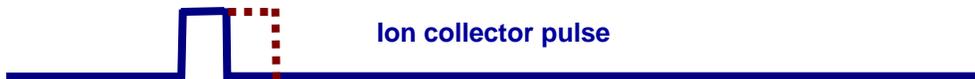
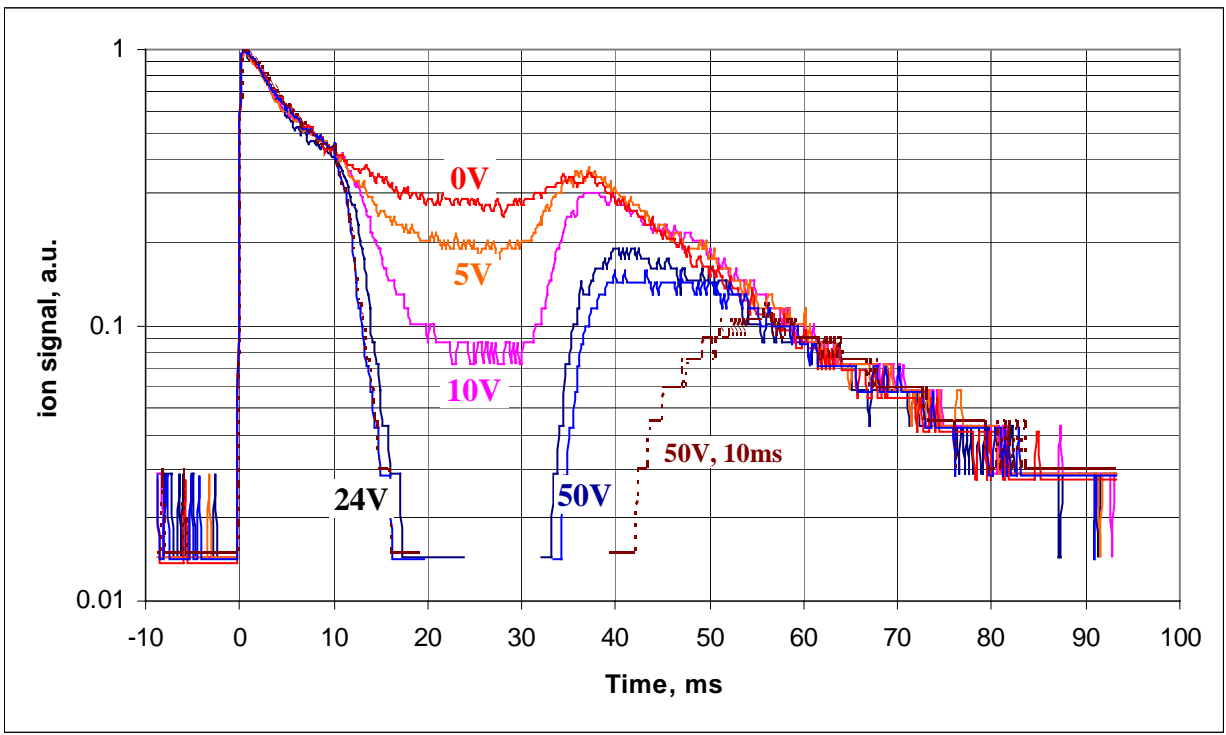

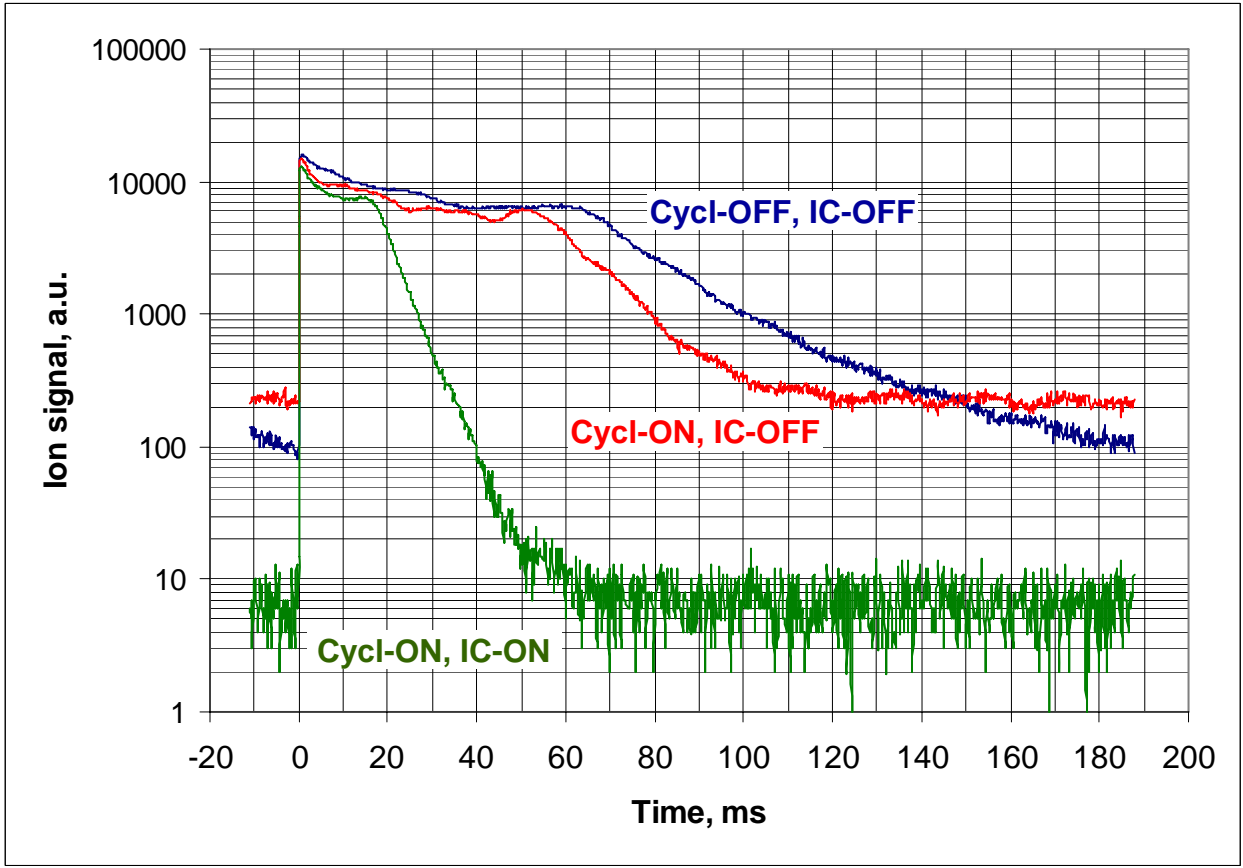

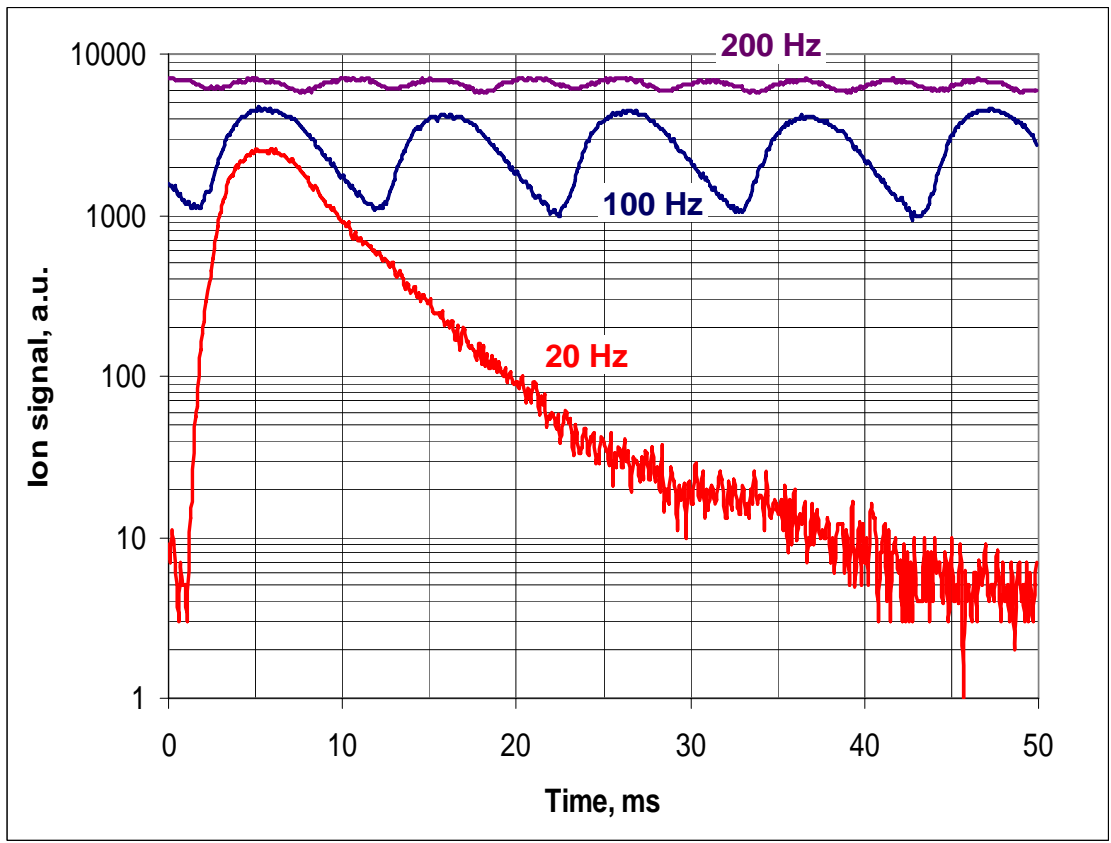

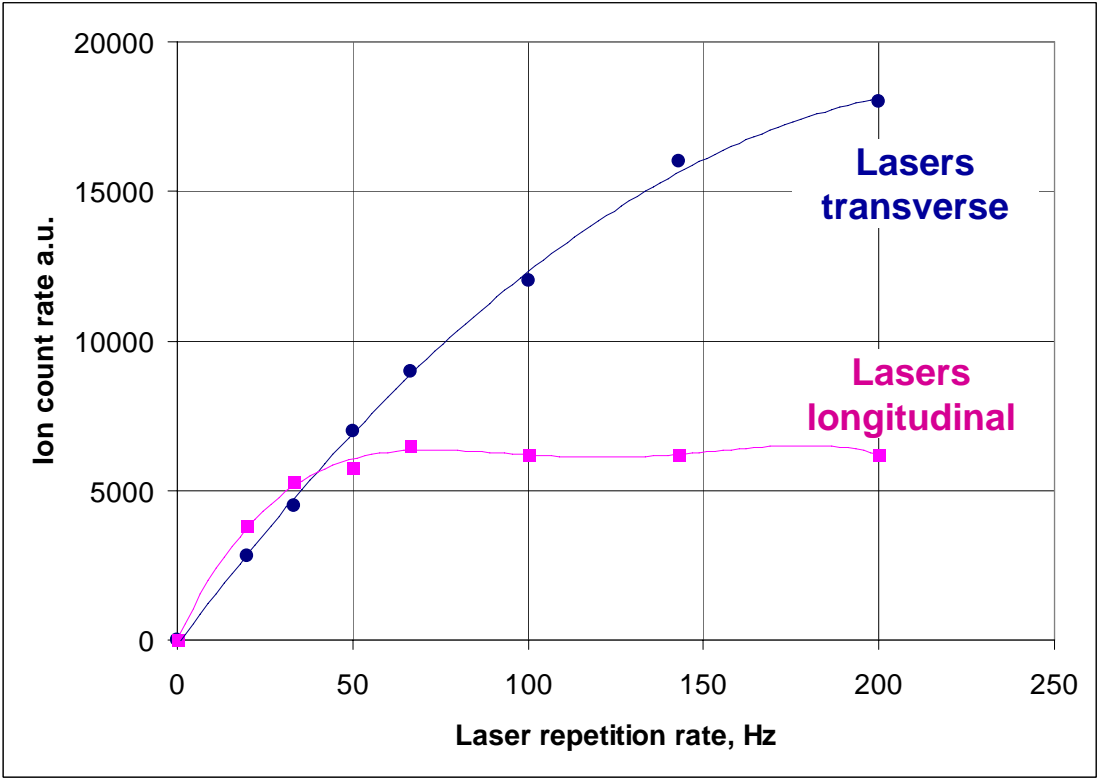

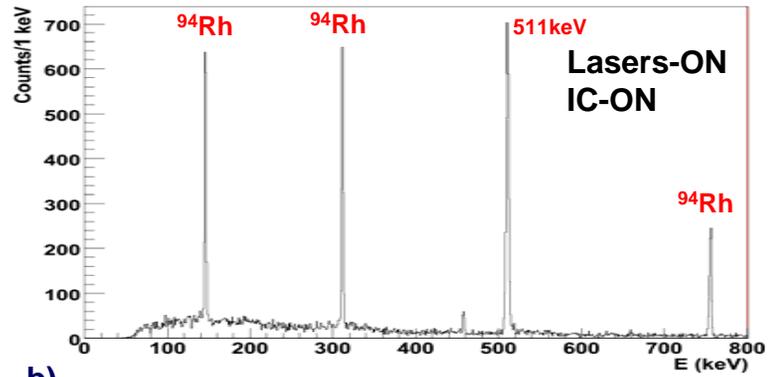

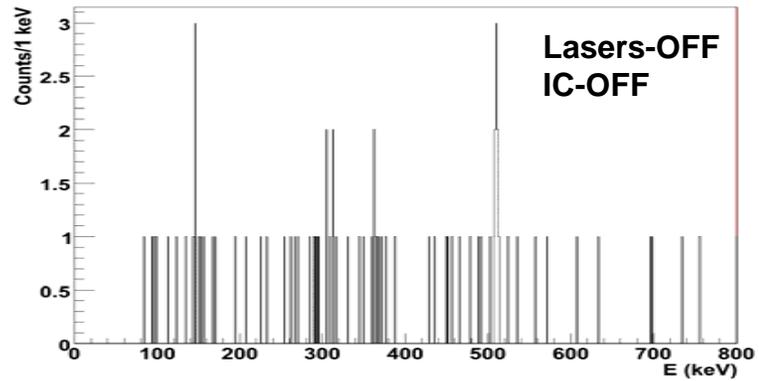

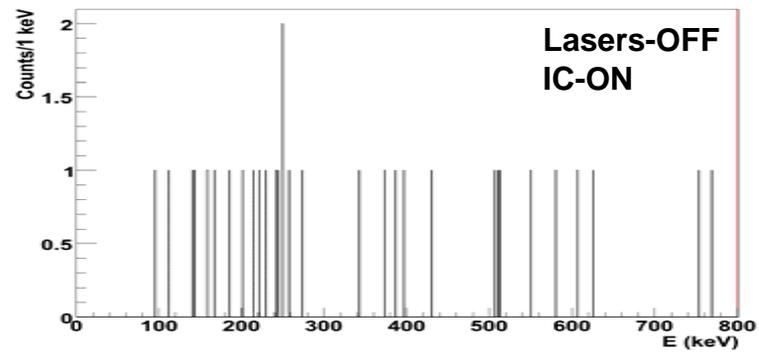

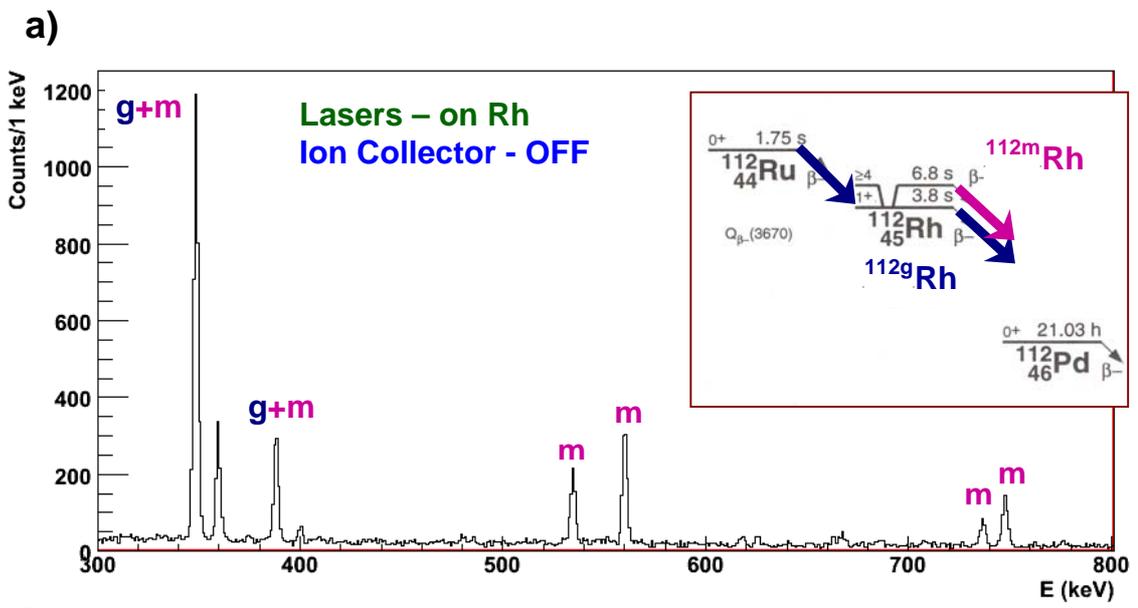
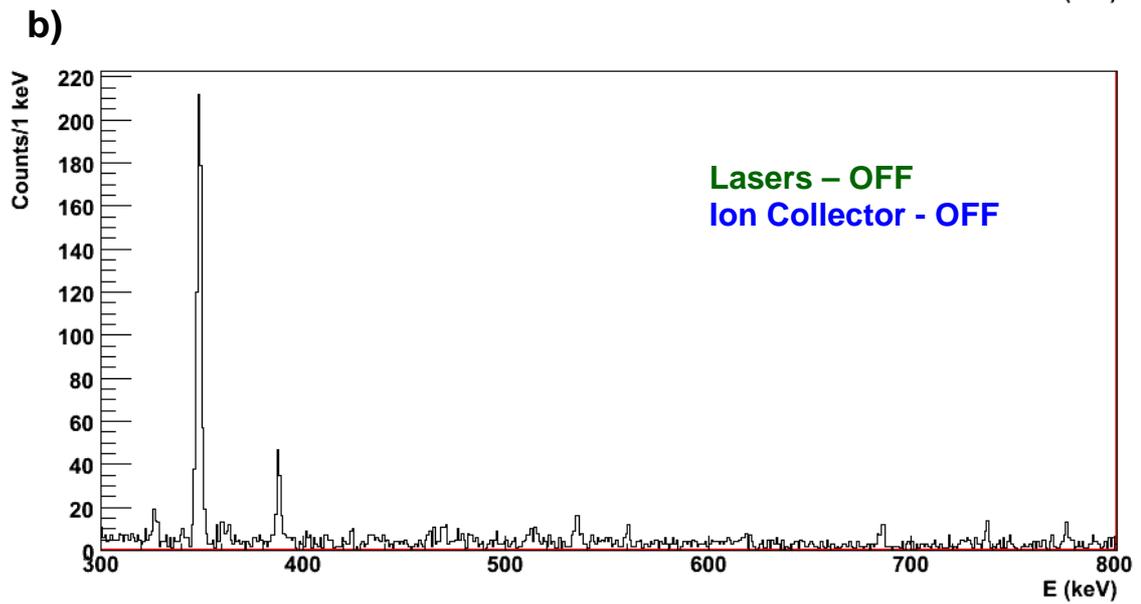
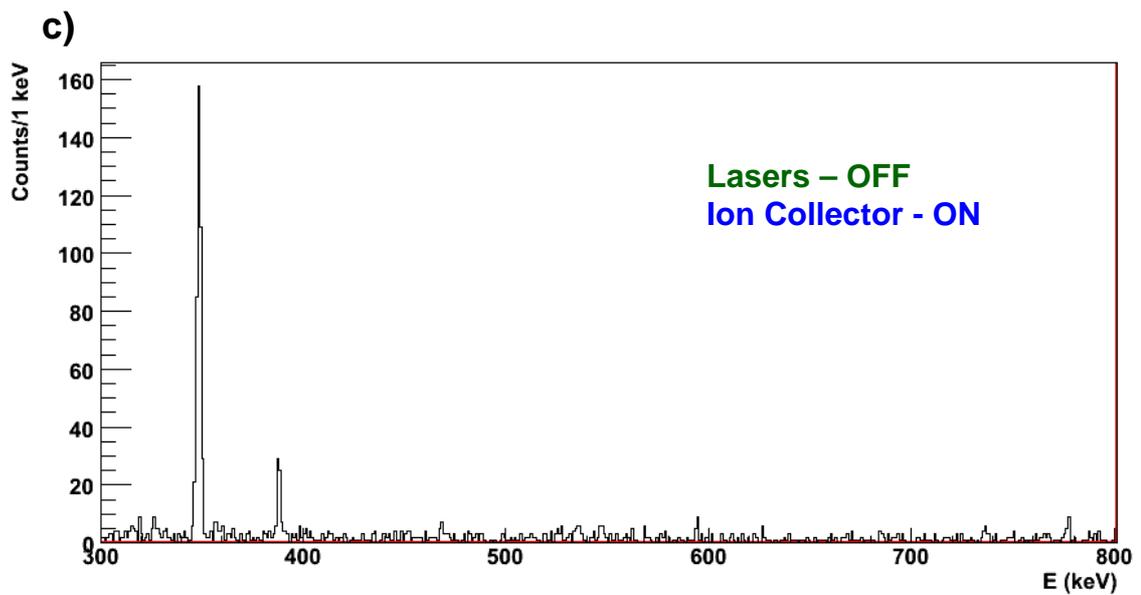

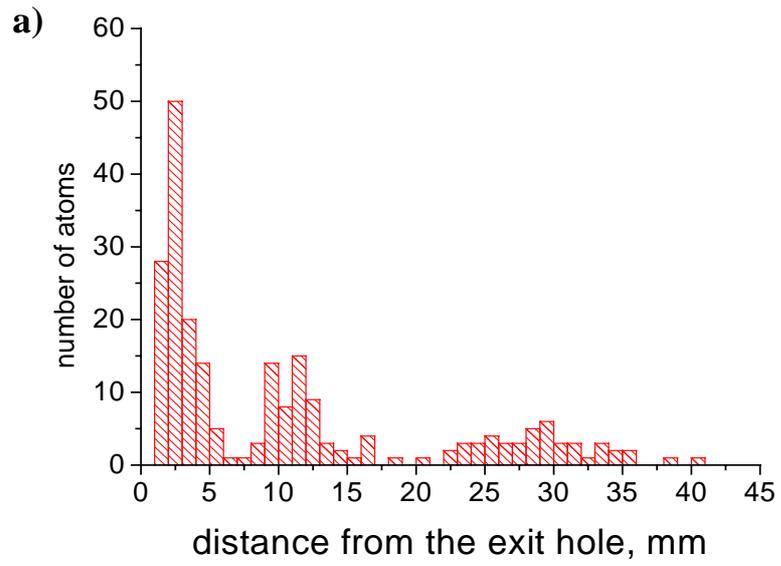
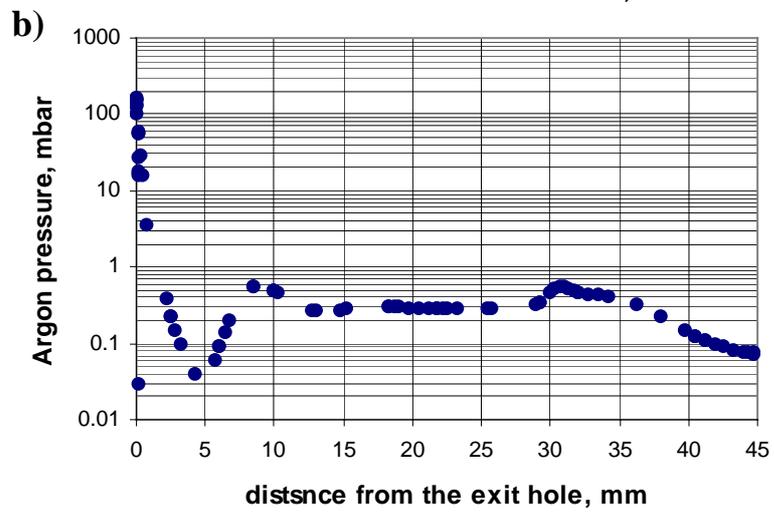

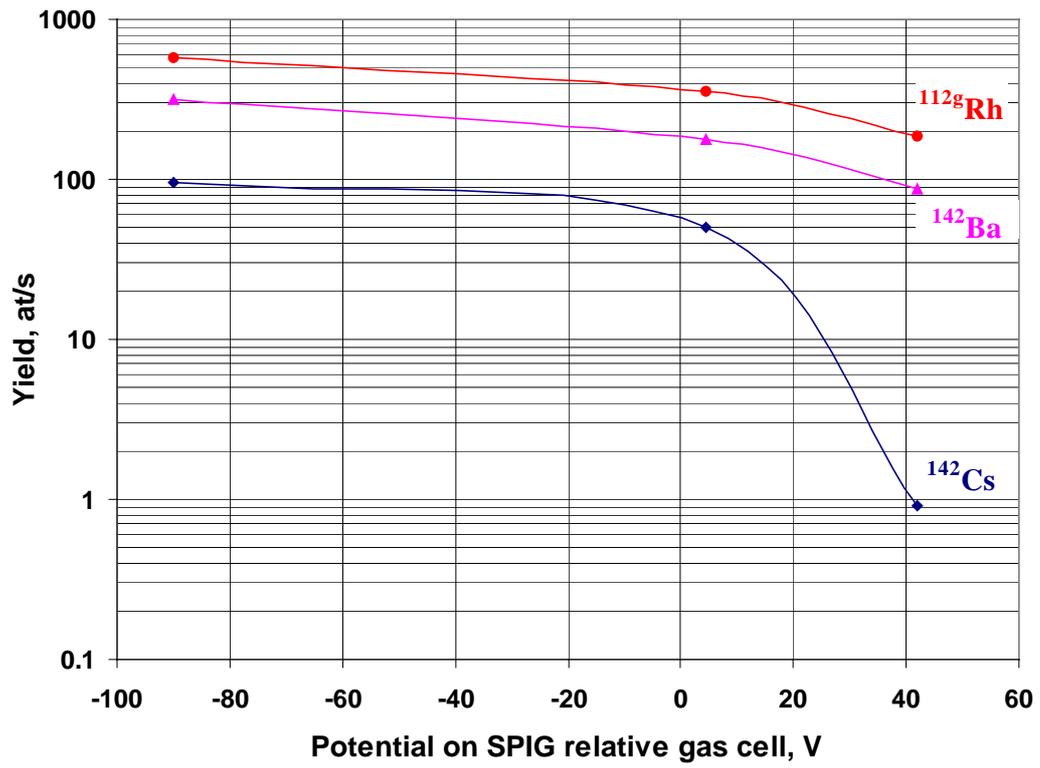